\documentclass[11pt]{article}
\usepackage{amsfonts}
\usepackage{amsmath, amssymb}
\usepackage[doublespacing]{setspace}
\usepackage[usenames,dvipsnames,svgnames,table]{xcolor}

\newtheorem{theorem}{Theorem}

\input{tcilatex}

\usepackage{hyperref}
\hypersetup{
    colorlinks=true,
    urlcolor=black,
    linkcolor=black,
    citecolor=black
}

\usepackage[
backend=biber,
style=apa,
sorting=nyt
]{biblatex}
\DeclareSourcemap{
  \maps[datatype=bibtex]{
    \map{
      \step[fieldset=issn, null]
      \step[fieldset=doi, null]
      \step[fieldset=url, null]
      \step[fieldset=urldate, null]
    }
  }
}

\newcommand{\citep}[2][]{\citeauthor{#2} (\citeyear[#1]{#2})}

\begin{document}

\title{\textbf{The unreasonable effectiveness of optimal transport in
economics}\\
\emph{submitted to the proceeding of the 2020 World Congress of the Econometric Society} }
\author{\textbf{Alfred Galichon}\thanks{%
New York University and Sciences Po, Email: alfred.galichon@nyu.edu. Support
from the European Research Council under Grant Agreement No. 866274
\textquotedblleft Equiprice\textquotedblright\, as well as research assistance by Gabriele Buontempo and Giovanni Montanari are gratefully acknowledged. }}
\maketitle

\emph{This paper is dedicated to the memory of Emmanuel Farhi (1978-2020).}

\section{Introduction}

The mathematical theory of optimal transport traces back to Monge in the
18th century, who asked the main questions, for which he provided deep
insights but left them unresolved. Regarded as a famous open problem
throughout the 19th century, it was revived, and finally solved, with the
advent of linear programming and works by Kantorovich, Koopmans, von
Neumann, Dantzig and others in the mid 20th century. While the theory partly
arose out of economic motivations (specifically resource allocation
problems), it soon drifted away from economics. A\ second revival has
occured since the 1990s, when insights from convex analysis were introduced
by Brenier, Rachev, and R\"{u}schendorf, and from geometry by Gangbo,
McCann, Villani and others.

In spite of this, up until recently, optimal transport has still been
reported missing from the standard toolbox of quantitative economics.
However, it turns out that many basic problems encountered in diverse
economic applications in various fields are \emph{optimal transport problems
in disguise}. Beyond intellectual curiosity, understanding this connection
is useful to make use of the mature set of results of optimal transport to
solve the problems, and in particular, to deal with questions of existence,
uniqueness, stability, and computation without reinventing the wheel.

This paper is a rapid overview of some of these connections, and some
extensions. It is admittedly skewed toward my own work, and borrows much
material from my \citeyear{galichon_optimal_2016} monograph, \emph{Optimal Transport Methods in Economics%
}, to which the reader is referred for details. I cover much of the material
from an empirical perspective every winter in the January edition of my
`math+econ+code' masterclasses (\url{www.math-econ-code.org}). In
mathematics, a useful read is Santambrogio's \emph{Optimal Transport for
Applied Mathematicians} \citeyear{santambrogio_optimal_2015}, or Villani's \citeyear{villani_topics_2003} introductory lecture notes 
\emph{Topics in Optimal Transportation}. For computational aspects, \citeauthor{peyre_computational_2019}'s \emph{Computational Optimal Transport} is a useful complement.
Villani's \citeyear{villani_optimal_2009} treatise \emph{Optimal Transport: Old and New} remains the
most exhaustive reference on the topic.

\section{Optimal transport\ in a nutshell}

\subsection{Optimal transport duality\label{par:ot-duality}}

Let us describe the optimal transport problem in the discrete case. Assume a
central planner needs to match a population of workers, each of whom is
characterized by their type $x\in \mathcal{X}$, with a population of firms,
each of whom with type $y\in \mathcal{Y}$. A match between a worker of type $%
x$ and a firm of type $y$ produces output $\Phi _{xy}$, called \emph{%
transport surplus}. The set of types $\mathcal{X}$ and $\mathcal{Y}$ are
finite, and the total number of workers and firms are identical. We denote by $%
\left( p_{x}\right) $ and $\left( q_{y}\right) $ the vectors of probability
distribution over $\mathcal{X}$ and $\mathcal{Y}$, thus normalizing the
total mass of workers and firms to one: $\sum_{x\in \mathcal{X}%
}p_{x}=\sum_{y\in \mathcal{Y}}q_{y}=1$.

The central planner's problem is to form a matching, and therefore, to
decide on the mass $\pi _{xy}$ of pairs $xy$ to form. In the sequel, we
shall call $\pi _{xy}$ the \emph{optimal transport plan}. This quantity must
match everyone, namely satisfy the double set of constraints that all
workers of each type $x$ are assigned, $\sum_{y\in \mathcal{Y}}\pi
_{xy}=p_{x}$, and that all firms of type $y$ are assigned, namely $%
\sum_{x\in \mathcal{X}}\pi _{xy}=q_{y}$. With these constraints in mind, the
workers shall maximize total output, which is $\sum_{x\in \mathcal{X},y\in 
\mathcal{Y}}\pi _{xy}\Phi _{xy}$ the sum of the pairwise output weighted by
the mass of each pair. This yields the problem 
\begin{align}
\max_{\pi \geq 0}& \sum_{x\in \mathcal{X},y\in \mathcal{Y}}\pi _{xy}\Phi
_{xy}  \label{primal-noregul} \\
s.t.& \left\{ 
\begin{array}{c}
\sum_{y\in \mathcal{Y}}\pi _{xy}=p_{x}~\forall x\in \mathcal{X} \\ 
\sum_{x\in \mathcal{X}}\pi _{xy}=q_{y}~\forall y\in \mathcal{Y}%
\end{array}%
\right.  \notag
\end{align}%
which is clearly a linear programming problem, which we shall call the \emph{%
primal problem}. Note that the set of $\pi $ satisfying the constraints is
clearly nonempty, as the \emph{random matching} obtained by $\pi
_{xy}=p_{x}q_{y}$ does satisfy the constraints. However, this matching is
not optimal in general.

It is a basic result in linear programming that in this case, the value of
the primal problem coincides with the value of the \emph{dual problem},
which is%
\begin{align}
\min_{u,v}& \sum_{x\in \mathcal{X}}p_{x}u_{x}+\sum_{y\in \mathcal{Y}%
}q_{y}v_{y}  \label{dual-noregul} \\
s.t.& u_{x}+v_{y}\geq \Phi _{xy}~\forall x\in \mathcal{X},y\in \mathcal{Y} 
\notag
\end{align}%
where the dual variables $u_{x}$ and $v_{y}$ are the Lagrange multipliers
respectively associated with the primal constraints $\sum_{y\in \mathcal{Y}%
}\pi _{xy}=p_{x}$ and $\sum_{x\in \mathcal{X}}\pi _{xy}=q_{y}$, while the
primal variables $\pi _{xy}\geq 0$ serve as Lagrange multipliers associated
with the dual constraints $u_{x}+v_{y}\geq \Phi _{xy}$.

Lastly, another important result in linear programming, \emph{complementary
slackness}, asserts that $\pi _{xy}>0\implies u_{x}+v_{y}=\Phi _{xy}$: if a
Lagrange multiplier is strictly positive, then the corresponding dual
constraint is saturated.

\bigskip

Optimal transport is a far-reaching generalization of the finite-dimensional
duality discussed above to the case when $\mathcal{X}$ and $\mathcal{Y}$ are
much richer sets; in particular the theory applies to the case when $%
\mathcal{X}$ and $\mathcal{Y}$ are finite-dimensional vector spaces, and we
will not need more for most of the economic applications we will discuss. In
that case, letting $P$ and $Q$ be probability distributions over $\mathcal{X}
$ and $\mathcal{Y}$, we shall define $\mathcal{M}\left( P,Q\right) $ as the
set of joint probability distributions over $\mathcal{X}\times \mathcal{Y}$
with margins $P$ and $Q$, which is the set of joint probability distributions 
$\pi $ such that if $\left( X,Y\right) \sim \pi $, where $\sim $ is
understood as \textquotedblleft distributed as,\textquotedblright\ then $%
X\sim P$ and $Y\sim Q$. In this more general setting, the primal problem~(\ref%
{primal-noregul}) extends to%
\begin{equation}
\max_{\pi \in \mathcal{M}\left( P,Q\right) }\int_{\mathcal{X}\times \mathcal{%
Y}}\Phi \left( x,y\right) d\pi \left( x,y\right)
\label{primal-noregul-continuous}
\end{equation}%
while its dual, extending~(\ref{dual-noregul}), is 
\begin{eqnarray}
\min_{u,v} &&\int_{\mathcal{X}}u\left( x\right) dP\left( x\right) +\int_{%
\mathcal{Y}}v\left( y\right) dQ\left( y\right)
\label{dual-noregul-contuinous} \\
s.t.~ &&u\left( x\right) +v\left( y\right) \geq \Phi \left( x,y\right) 
\notag
\end{eqnarray}

The \textbf{Monge-Kantorovich theorem} provides assumptions under which the
former duality results are preserved in more general settings, that is: (i)
there exist primal solutions $\left( \pi _{xy}\right) $; (ii) there is no
duality gap, that is, the value of the dual problem~(\ref{dual-noregul-contuinous})
coincides with the value of the primal~(\ref{primal-noregul-continuous}); and (iii) there exist dual solutions 
$\left( u_{x}\right) $ and $\left( v_{y}\right) $.

\subsection{Some variants}

\subsubsection{Entropy regularized Optimal Transport\label{par:regul-ot}}

To facilitate computation, consider the previous primal problem with an
entropic regularization in the objective function. Take $\sigma >0$ a
parameter that can be made arbitrarily small. The primal problem%
\begin{align}
\max_{\pi \geq 0}& \sum_{x,y}\pi _{xy}\Phi _{xy}-\sigma \sum_{x,y}\pi
_{xy}\ln \pi _{xy}  \label{primal-regul} \\
s.t.& \left\{ 
\begin{array}{c}
\sum_{y}\pi _{xy}=p_{x} \\ 
\sum_{x}\pi _{xy}=q_{y}%
\end{array}%
\right.  \notag
\end{align}%
has dual%
\begin{equation}
\min_{u,v}\left\{ \sum_{x\in \mathcal{X}}p_{x}u_{x}+\sum_{y\in \mathcal{Y}%
}q_{y}v_{y}+\sigma \sum_{x,y}\exp \left( \frac{\Phi _{xy}-u_{x}-v_{y}}{%
\sigma }\right) -\sigma \right\} ,  \label{dual-regul}
\end{equation}%
and the optimal $\pi $ in~(\ref{primal-regul}) and the optimal $\left(
u,v\right) $ in~(\ref{dual-regul}) are related by%
\begin{equation}
\pi _{xy}=\exp \left( \frac{\Phi _{xy}-u_{x}-v_{y}}{\sigma }\right) .
\label{smooth-slackness}
\end{equation}

Again, this problem has extension to the case where $\mathcal{X}$ and $%
\mathcal{Y}$ are no longer discrete sets: this is the theory of \emph{%
Bernstein-Schr\"{o}dinger systems}, surveyed in \citep{leonard_survey_2014}.
Recently, progresses have been made on the computation of this problem in
particular through \emph{coordinate descent}:

Starting at $t=0$ with an initial estimate of $v_{y}^{t}:$

\begin{itemize}
\item Compute $(u_{x}^{t+1})$ in order to minimize the objective function
in~(\ref{dual-regul}), while keeping $v=(v_{y}^{t})$ fixed.

\item Compute $(v_{y}^{t+1})$ in order to minimize the objective function
in~(\ref{dual-regul}), while keeping $u=(u_{x}^{t+1})$ fixed.

\item Iterate over $t$, until the update to the $u$'s and the $v$'s are
below tolerance.
\end{itemize}

It is easy to see that both steps are explicit and one full iteration of the
algorithm expresses as%
\begin{equation}
\left\{ 
\begin{array}{l}
u_{x}^{t+1}=\sigma \log (\frac{1}{p_{x}}\sum_{y\in \mathcal{Y}}\exp (\frac{%
\Phi _{xy}-v_{y}^{t}}{\sigma })) \\ 
v_{y}^{t+1}=\sigma \log (\frac{1}{q_{y}}\sum_{y\in \mathcal{Y}}\exp (\frac{%
\Phi _{xy}-u_{x}^{t+1}}{\sigma })).%
\end{array}%
\right.  \label{ipfp}
\end{equation}

This is the \emph{iterated proportional fitting algorithm }(IPFP), which has
been rediscovered under many names\footnote{See \citep{idel_review_2016} for a historical survey.}:
\textquotedblleft matrix scaling\textquotedblright , \textquotedblleft RAS
algorithm\textquotedblright , \textquotedblleft Sinkhorn-Knopp
algorithm\textquotedblright , \textquotedblleft Kruithof's
method\textquotedblright , \textquotedblleft Furness
procedure\textquotedblright , \textquotedblleft biproportional fitting
procedure\textquotedblright , \textquotedblleft Bregman's
procedure\textquotedblright . In economics, this algorithm has been proposed
at least twice, once by Berry, Levinsohn and Pakes \citeyear{berry_automobile_1995} under the name
\textquotedblleft contraction mapping algorithm,\textquotedblright\ and once
by Guimares-Portugal (\citeyear{guimaraes_simple_2010}) in the context of the gravity equation in trade.
We will see some of these connections below. This algorithm has been
successfully applied to machine learning, see Cuturi (\citeyear{cuturi_sinkhorn_2013}) and Peyr\'{e}
and Cuturi (\citeyear{peyre_computational_2019}). The rates of convergence of this algorithm are by now
well understood thanks to the Hilbert projective metric, see Franklin and
Lorenz (\citeyear{franklin_scaling_1989}), and more recently, thanks to the theory of Bregman
divergences, see L\'{e}ger (\citeyear{leger_gradient_2020}).

\subsubsection{Variant with unassigned agents}

A variant of the problem leaves the agents the possibility of remaining
unassigned. When the total mass of workers differs from that of the firms,
we still denote $p_{x}$ be the mass of workers of type $x$ (no longer
interpreted as a probability) and $q_{y}$ the mass of firms of type $y$, and
we allow for $\sum_{x}p_{x}\neq \sum_{y}q_{y}$ to hold. The constraint is
now that the total mass of matched workers of type $x$ should be no greater
than $p_{x}$, and that the total mass of matched firms of type $y$ should
not exceed $q_{y}$; and the total surplus is still $\sum_{xy}\pi _{xy}\Phi
_{xy}$; the primal problem is now 
\begin{align}
\max_{\pi \geq 0}& \sum_{x\in \mathcal{X},y\in \mathcal{Y}}\pi _{xy}\Phi
_{xy}  \label{primal-singles} \\
s.t.& \left\{ 
\begin{array}{c}
\sum_{y\in \mathcal{Y}}\pi _{xy}\leq p_{x}~\forall x\in \mathcal{X} \\ 
\sum_{x\in \mathcal{X}}\pi _{xy}\leq q_{y}~\forall y\in \mathcal{Y}%
\end{array}%
\right.  \notag
\end{align}%
while the dual problem becomes%
\begin{align}
\min_{u\geq 0,v\geq 0}& \sum_{x\in \mathcal{X}}p_{x}u_{x}+\sum_{y\in 
\mathcal{Y}}q_{y}v_{y}  \label{dual-singles} \\
s.t.& u_{x}+v_{y}\geq \Phi _{xy}~\forall x\in \mathcal{X},y\in \mathcal{Y} 
\notag
\end{align}

As we see, these formulations only slightly differ from~(\ref{primal-noregul}%
) and~(\ref{dual-noregul}) respectively: the constraints in the primal
switch from an equality to an inequality, while the variables in the dual
are now subject to nonnegativity constraints. Because this leaves the
possibility of agents to remained unmatched (unemployed in the labor market;
singles in the marriage market), these problems are sometimes more relevant
for economic modelling. This model is called the Becker-Shapley-Shubik
model, after Becker (\citeyear{becker_theory_1973}) and Shapley-Shubik (\citeyear{shapley_assignment_1971}).

\subsection{Inverse optimal transport problem}

Understanding the \textquotedblleft direct problem\textquotedblright\ of
optimal transport as determining the optimal transport plan $\pi _{xy}$ in~(%
\ref{primal-noregul}) or~(\ref{primal-regul}) based on the transport surplus 
$\Phi _{xy}$, as described above, we now turn to the \textquotedblleft
inverse problem\textquotedblright\ of optimal transport: how to determine
the transport surplus $\Phi _{xy}$ based on the observation of an optimal
transport plan $\hat{\pi}_{xy}$. More specifically, we specify%
\begin{equation}
\Phi _{xy}^{\lambda }=\sum_{k}\lambda _{k}\phi _{xy}^{k}.
\label{param-of-Phi}
\end{equation}

Setting $\theta =\left( \lambda _{k},u_{x},v_{y}\right) $ and%
\begin{equation}
\pi _{xy}^{\theta }=\exp \left( \Phi _{xy}^{\lambda }-u_{x}-v_{y}\right) ,
\label{ansatz}
\end{equation}%
the inverse optimal transport problem consists of seeking the parameter $%
\theta $ such that $\pi _{xy}^{\theta }$ has the same margins and moments
as $\hat{\pi}_{xy}$, that is 
\begin{equation}
\sum_{y\in \mathcal{Y}}\pi _{xy}^{\theta }=\sum_{y\in \mathcal{Y}}\hat{\pi}%
_{xy}=:p_{x},~\sum_{x\in \mathcal{X}}\pi _{xy}^{\theta }=\sum_{x\in \mathcal{%
X}}\hat{\pi}_{xy}=:q_{y},~\sum_{x,y}\pi _{xy}^{\theta }\phi
_{xy}^{k}=\sum_{x,y}\hat{\pi}_{xy}\phi _{xy}^{k}.  \label{moments}
\end{equation}

This question solved by the following convex optimization problem:

\begin{theorem}
The unique $\lambda $ satisfying conditions~(\ref{moments}) is unique
solution to%
\begin{equation}
\min_{u,v,\lambda }\left\{ \sum_{x\in \mathcal{X}}p_{x}u_{x}+\sum_{y\in 
\mathcal{Y}}q_{y}v_{y}+\sum_{x,y}\exp \left( \Phi _{xy}^{\lambda
}-u_{x}-v_{y}\right) -\sum_{xy}\hat{\pi}_{xy}\Phi _{xy}^{\lambda }\right\}
\label{PPML}
\end{equation}%
whose dual is%
\begin{eqnarray}
&&\max_{\pi \geq 0}\left\{ -\sum_{xy}\pi _{xy}\ln \pi _{xy}\right\}
\label{PPML-dual} \\
s.t.~ &&\sum_{y\in \mathcal{Y}}\pi _{xy}=p_{x}~\left[ u_{x}\right]
,~\sum_{x\in \mathcal{X}}\pi _{xy}=q_{y}~\left[ v_{y}\right] ,~  \notag \\
&&\sum_{x,y}\pi _{xy}\phi _{xy}^{k}=\sum_{x,y}\hat{\pi}_{xy}\phi _{xy}^{k}~%
\left[ \lambda _{k}\right] .  \notag
\end{eqnarray}
\end{theorem}

The problem of parametric estimation of $\lambda $ is therefore the problem
of a Poisson pseudo-maximum likelihood estimation, similar to the technique
employed in trade to estimate the gravity equation (Santos Silva and
Tenreyro, \citeyear{silva_log_2006}). Galichon and Salani\'{e} (\citeyear{galichon_cupids_2020}) formulated the initial
connection with the Choo-Siow (\citeyear{choo_who_2006}) matching model, in the variant with singles.
Dupuy and Galichon (\citeyear{dupuy_personality_2014})\ studies a continuous version of this model.
Dupuy, Galichon and Sun (\citeyear{dupuy_estimating_2019}) add a Lasso-type penalization to estimate $\lambda $
under sparsity constraint, while Carlier, Dupuy, Galichon and Sun (\citeyear{carlier_sista_2021}) offer an
algorithm called SISTA (Sinkhorn+Iterative Soft Thresholding Algorithm) to
compute efficently the regularized problem by alternating coordinate descent
steps (Sinkhorn steps) on the $u_{x}$'s and the $v_{y}$'s, with a proximal
gradient descent step.

\section{Optimal transport in economics, finance and statistics}

\subsection{Family economics\label{par:marriage}}

As first understood by Becker (\citeyear{becker_theory_1973})\ and Shapley-Shubik (\citeyear{shapley_assignment_1971}), the duality
in optimal transport can be thought as a powerful welfare theorem, providing
the equivalence between optimal matchings (in the sense of the problem of a
central planner), and stable matchings (in a sense to be specified). Becker
applied this insight in his pioneering analysis of the marriage market, and
we now describe his analysis.

Consider the \textquotedblleft marriage\textquotedblright\ problem of
heterosexual men and women who need to decide to match. Men are distributed
according to a a mass vector $\left( p_{x}\right) $, while women are
distributed according to a mass vector $\left( q_{y}\right) $, where the
total mass of men and women don't have to coincide. It is assumed that if $x$
and $y$ decide to match, they enjoy a joint utility $\Phi _{xy}$, which they
need to split among them. Any agent remaining unmatched gets a reservation
utility equal to zero.

A \emph{stable marriage} is a specification of a joint distribution $\left(
\pi _{xy}\right) \geq 0$ over $\mathcal{X}\times \mathcal{Y}$ as well as
payoffs vectors $u_{x}$ and $v_{y}$ such that 
\begin{equation}
\left\{ 
\begin{array}{l}
\sum_{y\in \mathcal{Y}}\pi _{xy}+\pi _{x0}=p_{x},~\sum_{x\in \mathcal{X}}\pi
_{xy}+\pi _{0y}=q_{y} \\ 
u_{x}+v_{y}\geq \Phi _{xy} \\ 
u_{x}\geq 0,~v_{y}\geq 0 \\ 
\pi _{xy}>0\implies u_{x}+v_{y}=\Phi _{xy} \\ 
\pi _{x0}>0\implies u_{x}=0,\pi _{0y}>0\implies v_{y}=0%
\end{array}%
\right.  \label{family-eq}
\end{equation}

The first set of conditions implies that all agents either participate
in the matching market, or remain unmatched. The next conditions, namely $%
u_{x}+v_{y}\geq \Phi _{xy}$, imply that there is no blocking pair: if $%
u_{x}+v_{y}$ were less than $\Phi _{xy}$, then $x$ and $y$ would have an
incentive to quit their existing assignments and form a blocking pair, and
each achieve strictly greater utility than $u_{x}$ and $v_{y}$ respectively.
Similarly, $u_{x}\geq 0$ and $v_{y}\geq 0$ indicate that no one can achieve
an outcome worse than the reservation utility.

Finally, the last set of conditions expresses that if pairs $xy$ are
actually formed, then there must be a way to split the joint surplus $\Phi
_{xy}$ in such a way that $u_{x}$ and $v_{y}$ sum to $\Phi _{xy}$, while if
a positive mass of either $x$ or $y$ remain unmatched at equilibrium, the
payoff of the corresponding type should be zero.

It is not hard to see that equations~(\ref{family-eq}) are the complementary
slackness conditions associated with linear programming problem (\ref%
{primal-singles})-(\ref{dual-singles}). Hence:

\begin{theorem}[Becker-Shapley-Shubik]
$\left( \pi ,u,v\right) $ is a stable marriage in the sense of~(\ref%
{family-eq}) if and only if $\pi $ is an optimal solution to~(\ref%
{primal-singles}), and $\left( u,v\right) $ is an optimal solution to~(\ref%
{dual-singles}).
\end{theorem}

This linear programming formulation is especially attractive for
computational purposes, see chapter 3.4 of Galichon (\citeyear{galichon_optimal_2016}).

\subsection{Labor economics\label{par:labor}}

In a realistic model of the labor market, not all jobs offering the same wage
are as attractive for the workers. Hence, we need to capture the job
amenities as the monetary valuations for working certain type of
jobs conditional on being a certain type of worker. Let $\alpha _{xy}$ be the
monetary valuation of employer $y$'s amenities for worker $x$, and let $%
\gamma _{xy}$ be the monetary output of worker $x$ working for employer $y$.
As before, we normalize to zero the payoff of unassigned agents.

Let $w_{xy}$ be the wage that $x$ receives if working for $y$, which is
determined at equilibrium. The worker and the firm problems are respectively%
\begin{equation}
u_{x}=\max_{y\in \mathcal{Y}}\left\{ \alpha _{xy}+w_{xy},0\right\} \text{
and }v_{y}=\max_{x\in \mathcal{X}}\left\{ \gamma _{xy}-w_{xy},0\right\}
\label{indirect-utils-becker}
\end{equation}%
from which it follows that, defining the total output associated with an $xy$
match as the sum of monetary amenity plus production, namely $\Phi
_{xy}=\alpha _{xy}+\gamma _{xy}$, an equilibrium on the labor market should
be such that $\left( \pi ,u,v\right) $ should be a stable matching in the
sense of~(\ref{family-eq}). Once $\left( u,v\right) $, which is solution to~(%
\ref{dual-singles}) has been computed, one can compute the vector of
equilibrium wages $w_{xy}$ by%
\begin{equation}
\gamma _{xy}-v_{y}\leq w_{xy}\leq u_{x}-\alpha _{xy}.  \label{becker-ineq}
\end{equation}%
Note that for pairs $xy$ that are actually formed at equilibirum, $\pi
_{xy}>0$ implies that $u_{x}+v_{y}=\Phi _{xy}$, and thus the upper bound $%
u_{x}-\alpha _{xy}$ coincides with the lower bound $\gamma _{xy}-v_{y}$. For
other pairs, the upper bound may differ from the lower bound, which is a
typical situation in equilibrium, where the price vectors need not be unique
outside of the equilibrium path.

\subsection{Trade\label{par:trade}}

The structural gravity equations in international trade, introduced by
Anderson (\citeyear{anderson_gravity_2003}), with antecedents in Alan Wilson (\citeyear{wilson1969use}), has been described as a
\textquotedblleft workhorse\textquotedblright\ model in that field (Head and
Mayer, \citeyear{head_gravity_2013}). Letting $\mathcal{X}$ be the set of countries, we define $\hat{%
\pi}_{xy}$ as the observed trade flow from country $x\in \mathcal{X}$ to
country $y\in \mathcal{X}$. Letting $p_{x}=\sum_{y\neq x}\hat{\pi}_{xy}$ be
the total volume of country $x$'s exports, and $q_{y}=\sum_{x\neq y}\hat{\pi}%
_{xy}$ be the total volume of country $y$'s imports, the gravity model
assumes that 
\begin{equation}
\pi _{xy}^{\lambda ,u,v}=\exp \left( \Phi _{xy}^{\lambda }-u_{x}-v_{y}\right)
\label{gravity-eq}
\end{equation}%
where $\Phi _{xy}^{\lambda }=\sum_{k}\phi _{xy}^{k}\lambda _{k}$ and the $%
\phi _{xy}^{k}$'s are various measures of proximity between country $x$ and
country $y$. The exporter and importer fixed effects $u_{x}$ and $v_{y}$ are
called \textquotedblleft multilateral resistances\textquotedblright\ and are
adjusted by fitting the total imports and exports%
\begin{equation}
\left\{ 
\begin{array}{c}
p_{x}=\sum_{y\neq x}\exp \left( \Phi _{xy}^{\lambda }-u_{x}-v_{y}\right) \\ 
q_{y}=\sum_{x\neq y}\exp \left( \Phi _{xy}^{\lambda }-u_{x}-v_{y}\right)%
\end{array}%
\right. .  \label{balance-of-trade}
\end{equation}

As understood by Wilson (\citeyear{wilson1969use}), \ $\pi ^{\lambda ,u,v}$ is the solution to
the regularized optimal transport problem~(\ref{primal-regul}), while $u_{x}$
and $v_{y}$ are solution to its dual~(\ref{dual-regul}). Moreover, $\theta
=\left( \lambda ,u,v\right) $ can estimated as an inverse optimal transport
problem~(\ref{PPML}), as suggested by the influencial paper of Santos Silva
and Tenreyro (\citeyear{silva_log_2006}), who connect the procedure with a Poisson regression.
The link with inverse optimal transport and matching problems is made in
Dupuy, Galichon and Sun (\citeyear{dupuy_estimating_2019}).

\subsection{Hedonic models\label{par:hedonic}}

Consider a quasilinear hedonic model where each producer $x\in \mathcal{X}$
produces one unit of good and chooses in which quality $z\in \mathcal{Z}$.
Each consumer $y\in \mathcal{Y}$ consumes one unit of good, and chooses in
which quality $z\in \mathcal{Z}$. The mass of the producers and consumers
are respectively distributed according to vectors $\left( p_{x}\right) $ and 
$\left( q_{y}\right) $. There is a price $P_{z}$, determined at equilibrium,
for one unit of the good in quality $z$, and a producer of type $x$ incurs a
profit $P_{z}-C_{xz}$ of producing quality $z$ at that price where $C$ is a
cost, while a consumer of type $y$ derives a utility $U_{yz}-P_{z}$ of
consuming utility $z$ at that price. Both producers and consumers can opt
out of the market and get profit or utility zero in that case.

In a hedonic equilibrium (Ekeland, Heckman and Nesheim, \citeyear{ekeland_identification_2004}), demand and
supply are formed by the producer's and consumer's problems which are
respectively 
\begin{equation}
u_{x}=\max_{z\in \mathcal{Z}}\left\{ P_{z}-C_{xz},0\right\} \text{ and }%
v_{y}=\max_{z\in \mathcal{Z}}\left\{ U_{yz}-P_{z},0\right\} .
\label{indirect-utilities-hedonic}
\end{equation}

Chiappori, McCann and\ Nesheim (\citeyear{chiappori_hedonic_2010}) have shown that this problem is
actually an optimal transport problem of the type~(\ref{primal-singles})
between consumers and producers, with a matching surplus equal to%
\begin{equation}
\Phi _{xy}=\max_{z}\left\{ U_{yz}-C_{xz}\right\}
\label{joint-surplus-hedonic}
\end{equation}%
and the indirect utilities $u_{x}$ and $v_{y}$ are determined by~(\ref%
{dual-singles}). The intuition for the result is limpid: if $x$ and $y$
decide to exchange a good, they should pick the good which is cost efficient
in the sense that it maximizes their total joint surplus. The price vector $%
P_{z}$ will be deduced from $u_{x}$ and $v_{y}$ by the set of inequalities 
\begin{equation}
\min_{x\in \mathcal{X}}\left\{ u_{x}+C_{xz}\right\} \geq P_{z}\geq
\max_{y\in \mathcal{Y}}\left\{ U_{yz}-v_{y}\right\}  \label{hedonic-ineq}
\end{equation}%
where -- similarly to the wage determination in equation~\eqref{becker-ineq}
-- the lower bound and the upper bound will coincide as soon as the quality $%
z$ is actually traded at equilibrium.

\subsection{Discrete choice models\label{par:discrete-choice}}

Recently, an intimate connection between optimal transport theory and
discrete choice models has been explored, which we now describe. Consider
the (additive) discrete choice problem where a consumer $i$ drawn from a
population faces a choice between a finite set of alternatives $y\in 
\mathcal{Y}$. Consumer $i$'s problem is 
\begin{equation}
u\left( \varepsilon _{i}\right) =\max_{y\in \mathcal{Y}}\left\{
V_{y}+\varepsilon _{iy}\right\}  \label{dc-model}
\end{equation}%
where $V_{y}$ is the systematic utility that every consumers associate with
alternative $y$, and $\left( \varepsilon _{iy}\right) _{y\in \mathcal{Y}}$
is drawn from a random vector over $\mathbb{R}^{\mathcal{Y}}$ with
distribution $\mathbf{P}$, which is assumed to have a density. The
distribution of the random part of the utility $\varepsilon $ induces a
choice probability, or market share $Q_{y}\left( V\right) $ which is the
probability that $y$ is chosen by a consumer $i$ drawn from the population,
formally expresses as\footnote{%
Note that as $\varepsilon $ has a density, the probability of ties is zero,
and therefore the $\arg \max $ has almost surely one element.}%
\begin{equation}
Q_{y}\left( V\right) =\Pr \left( y\in \arg \max_{y\in \mathcal{Y}}\left\{
V_{y}+\varepsilon _{iy}\right\} \right) .  \label{demand-map}
\end{equation}

The \emph{demand inversion problem}, popularized by Berry (\citeyear{berry_estimating_1994}) and Berry,
Levinsohn and Pakes (\citeyear{berry_automobile_1995}, hereafter BLP) consists of, given a vector of
market shares $q_{y}$, how to look for a vector of systematic utility $V$
such that $Q\left( V\right) =q$. This problem is a key step in BLP's
estimation procedure, which consists of computing $V$ by demand inversion,
and then running an instrumental variable regression on $V$.

\bigskip

Galichon and Salani\'{e} (\citeyear{galichon_cupids_2020}) showed that the problem of discrete choice
inversion is, in fact, isomorphic to an optimal transport problem.

\begin{theorem}[Galichon-Salani\'{e}, part 1]
\label{thm:gal-sal1}The following statements are equivalent:

(i) $Q\left( V\right) =q$, that is $V$ is the solution to inversion problem
of the discrete choice model in~(\ref{dc-model}), and

(ii) There exist $\left( u,v\right) $ with $v=-V$ such that $\left(
u,v\right) $ is solution to the dual optimal transport problem with surplus $%
\Phi \left( \varepsilon ,y\right) :=\varepsilon _{y}$%
\begin{align}
\min_{u,v}& \int_{\varepsilon \in \mathbb{R}^{\mathcal{Y}}}u\left(
\varepsilon \right) d\mathbf{P}\left( \varepsilon \right) +\sum_{y\in 
\mathcal{Y}}q_{y}v_{y}  \label{demand-inversion-OT} \\
s.t.~& u\left( \varepsilon \right) +v_{y}\geq \varepsilon _{y}\text{ }%
\forall \varepsilon \in \mathbb{R}^{\mathcal{Y}},\forall y\in \mathcal{Y}. 
\notag
\end{align}
\end{theorem}

This result was extended to the nonsmooth case (where no regularity
assumption is made on the distribution of $\varepsilon $) by Chiong,
Galichon and Shum (\citeyear{chiong_duality_2016}), where a linear programming approch was provided
for computational purposes. It has been extended to the continuous choice by
Chernozhukov, Galichon, Henry and Pass (\citeyear{chernozhukov_identification_2021}), and beyond additive random
utility models by Bonnet et al. (\citeyear{bonnet_yogurts_2015}).

A philosophical consequence of theorem~\ref{thm:gal-sal1} is that -- at
least from a mathematical standpoint -- there is no relevant distinction
between \textquotedblleft one-sided\textquotedblright\ and \textquotedblleft
two-sided\textquotedblright models. We think of a discrete choice problem
as a situation where conscient creatures called \textquotedblleft
consumers\textquotedblright\ choose inanimate objects called \textquotedblleft
yogurts\textquotedblright . However, the equivalence described in theorem~%
\ref{thm:gal-sal1} shows that this situation is mathematically equivalent to
a situation where consumers and yogurts would match, which is itself fully equivalent to a 
situation where yogurts choose consumers! This is a manifestation of Coase's principle: no matter how the
utility is initially distributed, that is, no matter if consumers have
preferences for yogurts or if yogurts have preferences for consumers, a
Pareto efficient outcome should be reached in any case, and the bargaining
process, here the yogurt price adjustment, allows to implement this outcome.

\bigskip

Interestingly, theorem~\ref{thm:gal-sal1} can be extended to mixed logit
models, such as BLP's random coefficient logit model. Consider now a variant 
\begin{equation}
u\left( \varepsilon _{i}\right) =\max_{y\in \mathcal{Y}}\left\{
V_{y}+\varepsilon _{iy}+\sigma \eta _{y}\right\}  \label{mixed-logit-model}
\end{equation}%
where $\left( \varepsilon _{y}\right) \sim \mathbf{P}$ as before, while $%
\left( \eta _{y}\right) $ is a vector of i.i.d. random variables with a
Gumbel distribution, independent from $\left( \varepsilon _{y}\right) $. Let 
$Q_{y}^{\sigma }\left( V\right) $ be the corresponding market share defined
for each entry $y\in \mathcal{Y}$.

\begin{theorem}[Galichon-Salani\'{e}, part 2]
\label{thm:gal-sal2}The following statements are equivalent:

(i) $Q^{\sigma }\left( V\right) =q$, that is $V$ is the solution to
inversion problem of the discrete choice model in~(\ref{mixed-logit-model}),
and

(ii) There exist $\left( u,v\right) $ with $v=-V$ such that $\left(
u,v\right) $ is solution to the dual regularized optimal transport problem
with surplus $\Phi \left( \varepsilon ,y\right) :=\varepsilon _{y}$%
\begin{equation}
\min_{u,v}\int_{\varepsilon \in \mathbb{R}^{\mathcal{Y}}}u\left( \varepsilon
\right) d\mathbf{P}\left( \varepsilon \right) +\sum_{y\in \mathcal{Y}%
}q_{y}v_{y}+\sigma \sum_{y\in \mathcal{Y}}\int_{\varepsilon \in \mathbb{R}^{%
\mathcal{Y}}}\exp \left( \frac{\varepsilon _{y}-u\left( \varepsilon \right)
-v_{y}}{\sigma }\right) d\varepsilon .  \label{BLP-regulOT}
\end{equation}
\end{theorem}

Note that~(\ref{BLP-regulOT}) is the same problem as~(\ref{dual-regul})
where the summation on $\varepsilon $ has been replaced by a continuous
integrals; however, in the sample version, we considering a sample $%
\varepsilon _{1},...,\varepsilon _{N}\,$\ from distribution $\mathbf{P}$,
and the integrals are replaced by sums.

As shown in Bonnet et al. (\citeyear{bonnet_yogurts_2015}), the coordinate descent algorithm described
in paragraph~\ref{par:regul-ot} coincides with BLP's celebrated
\textquotedblleft contraction mapping algorithm.\textquotedblright\ This
observation led the former authors to propose a demand inversion procedure
that extends to the non-additive case.

\subsection{Derivative pricing}

Consider two stocks, and let $X$ and $Y$ be random variables standing for
the value of these stocks at a horizon of time in the future. The
fundamental theorem of asset pricing (see Duffie \citeyear{duffie_dynamic_2001}) asserts that if there
is a complete market of options with $X$ as an underlying, then there is a
distribution $\mathbf{P}$ called \emph{martingale measure} such that the
price of an option whose payoff is $u\left( X\right) $ shall be $\mathbb{E}_{%
\mathbf{P}}\left[ u\left( X\right) \right] $. We shall assume that this is
the case, and that there is a martingale measure $\mathbf{Q}$ such that the
price of any option with payoff $v\left( Y\right) $ is $\mathbb{E}_{\mathbf{Q%
}}\left[ v\left( Y\right) \right] $.

However, we shall not assume that there is a complete market of options on
the joint realization of the underlying pair $\left( X,Y\right) $, hence we
cannot infer a joint martingale measure $\pi \left( x,y\right) $ based on
the quoted prices. For a trader wishing to introduce a new option on the
pair $\left( X,Y\right) $, some restrictions must however be considered; in
particular, if the option's payoff is of the form $a\left( X\right) +b\left(
Y\right) $, its price must be $\mathbb{E}_{\mathbf{P}}\left[ a\left(
X\right) \right] +\mathbb{E}_{\mathbf{Q}}\left[ b\left( Y\right) \right] $,
otherwise the trader would face an arbitrage opportunity. But in general,
the price of an option with a payoff $\Phi \left( X,Y\right) $ that is not
additively separable cannot exceed%
\begin{equation}
\max_{\pi \in \mathcal{M}\left( P,Q\right) }\mathbb{E}_{\pi }\left[ \Phi
\left( X,Y\right) \right] .  \label{opt-bound-static}
\end{equation}

The Monge-Kantorovich duality will give us \emph{sharp arbitrage bounds} for
the price of this option, and will provide arbitrage strategies, as
explained in Galichon, Henry-Labord\`{e}re and Touzi (\citeyear{galichon_stochastic_2014}):

\begin{theorem}
An option whose payoff $\Phi \left( X,Y\right) $ is priced at $V$ is not
subject to an arbitrage opportunity based on the two single-underlying
option markets if and only if 
\begin{equation}
\underset{s.t.~u\left( x\right) +v\left( y\right) \leq \Phi \left(
x,y\right) }{\max_{u,v}\mathbb{E}_{\mathbf{P}}\left[ u\left( X\right) \right]
+\mathbb{E}_{\mathbf{Q}}\left[ v\left( Y\right) \right] }\leq V\leq \underset%
{s.t.~u\left( x\right) +v\left( y\right) \geq \Phi \left( x,y\right) }{%
\min_{u,v}\mathbb{E}_{\mathbf{P}}\left[ u\left( X\right) \right] +\mathbb{E}%
_{\mathbf{Q}}\left[ v\left( Y\right) \right] }  \label{option-bounds}
\end{equation}
\end{theorem}

In other words, the price of the option should be bounded above by the price
of the \emph{cheapest overreplicating portfolio}, while it should be bounded
below by the price of the \emph{costliest underreplicating portfolio}.

The above discussion has assumed that the pair of underlyings $X$ and $Y$
were the realizations of two assets prices at the same time. However, some
derivatives are written on the same underlying asset at two different dates
in the future. Assume that $X$ is the value of a stock at a future date, and 
$Y$ is the stock value at a later date. We then have an additional
restriction, which is that in any martingale measure, $\mathbb{E}_{\pi }%
\left[ Y|X\right] =X$ expresses absence of arbitrage. The option bound
problem~(\ref{opt-bound-static}) now becomes 
\begin{eqnarray}
\max_{\pi \in \mathcal{M}\left( P,Q\right) } &&\mathbb{E}_{\pi }\left[ \Phi
\left( X,Y\right) \right]  \label{opt-bound-dynamic} \\
s.t.~ &&\mathbb{E}_{\pi }\left[ Y|X\right] =X  \notag
\end{eqnarray}%
for which the Monge-Kantorovich duality extends and interprets as
incorporating dynamic arbitrage strategies; see an exposition from a
financial engineering's point of view in Pierre Henry-Labord\`{e}re (\citeyear{henry-labordere_model-free_nodate})'s
insightful book.

\subsection{Quantiles}

There is an intimate connection between optimal transport and the notion of
quantile. Consider the optimal transport problem described in~(\ref%
{primal-noregul-continuous}) with $\mathcal{X}=\mathcal{Y}=\mathbb{R}$, $P=%
\mathcal{U}\left( \left[ 0,1\right] \right) $ the uniform distribution on
the unit interval, $Q$ a distribution with finite second moments, and $\Phi
\left( x,y\right) =xy$.

Then, as explained in chapter 4 of Galichon (\citeyear{galichon_optimal_2016}), the solution $\left(
X,Y\right) \sim \pi $ to problem~(\ref{primal-noregul-continuous}) is a
random pair such that $Y=F_{Q}^{-1}\left( X\right) $. Further, the solution $%
\left( u,v\right) $ to problem~(\ref{dual-noregul-contuinous}) is such that $%
u^{\prime }\left( x\right) =F_{Q}^{-1}\left( x\right) $ and $v^{\prime
}\left( y\right) =F_{Q}\left( y\right) $. Hence the primal solution involves
the quantile transform, and the dual solutions are simply primitives of the
quantile map and the cumulative distribution function.

\subsubsection{Multivariate quantiles}

This connection led to the definition of a notion of \textbf{multivariate
quantiles}: when $Y$ is multivariate, say has $d$ dimensions, one can extend
the above setting to $X\sim P=\mathcal{U}\left( \left[ 0,1\right]
^{d}\right) $ and to $\Phi \left( x,y\right) =x^{\top }y$ and, if $\left(
u,v\right) $ is a solution to problem~(\ref{dual-noregul-contuinous}) in
that case, the map $x\rightarrow \nabla u\left( x\right) $ is defined as the
multivariate quantile associated with distribution $Q$. By Brenier's theorem
(Brenier \citeyear{brenier_polar_1987}), $\nabla u\left( X\right) $ has distribution $Q$,
generalizing the well-known fact in the univariate case that the quantile
map associated with a distribution pushes the uniform distribution on the
unit interval onto the distribution. This new notion of multivariate
quantiles found applications to risk measures (Ekeland, Galichon and Henry,
\citeyear{ekeland_comonotonic_2012}), decision theory (Galichon and Henry, \citeyear{galichon_dual_2012}), and multivariate depth
(Hallin, Chernozhukov, Galichon and Henry, \citeyear{chernozhukov_mongekantorovich_2017}).

\subsubsection{Quantile regression}\label{par:quantile-regression}

There is an intimate connection between optimal transport and quantile
regression, that is explored in a series of paper by Carlier, Chernozhukov and
Galichon (\citeyear{carlier_vector_2016}, \citeyear{carlier_vector_2017}) and Carlier, Chernozhukov, De Bie and Galichon (\citeyear{carlier_vector_2020}).
We follow the latter paper in the present exposition. Quantile regression
(see Koenker, \citeyear{koenker_quantile_2005}) attempts to fit a parametric dependence of the
conditional $\tau $-th quantile of a random variable $Y$ conditional on the
value of $X$, a random vector on $\mathbb{R}^{k}$, as%
\begin{equation}
Q_{Y|X}\left( \tau |x\right) =\beta \left( \tau \right) ^{\top }x.
\label{quantile-spec}
\end{equation}%
where $\beta \left( \tau \right) \in \mathbb{R}^{k}$ is the parameter of
interest, defined for each value of $\tau \in \left[ 0,1\right] $. Since
Koenker and Bassett (\citeyear{koenker_regression_1978}), this problem has been recognized as a convex
optimization problem in the population%
\begin{equation}
\min_{\beta \left( \tau \right) \in \mathbb{R}^{k}}\mathbb{E}_{\mathbf{P}}%
\left[ \rho _{\tau }\left( Y-\beta \left( \tau \right) ^{\top }X\right) %
\right]  \label{qreg-pop}
\end{equation}%
where the loss function $\rho _{\tau }\left( z\right) =\tau z^{+}+\left(
1-\tau \right) z^{-}$, and $\mathbf{P}$ denotes the joint distribution of $%
\left( X,Y\right) $. The sample analog of~(\ref{qreg-pop}) is a linear
programming problem, yielding to a simple and computationally efficient
estimation of $\beta $. The full curve $\tau \rightarrow \beta \left( \tau
\right) $ can be estimated by summation of the objective functions in~(\ref%
{qreg-pop}) over $\tau \in \left[ 0,1\right] $, yielding%
\begin{equation}
\min_{\beta \in \mathbb{R}^{k\times \left[ 0,1\right] }}\mathbb{E}_{\mathbf{P%
}}\left[ \int_{0}^{1}\rho _{\tau }\left( Y-\beta \left( \tau \right) ^{\top
}X\right) d\tau \right] .  \label{qreg-pop-integrated}
\end{equation}

When specification~(\ref{quantile-spec}) is correct, the map $\tau
\rightarrow \beta \left( \tau \right) ^{\top }x$ which is picked up is an
actual quantile, and therefore nondecreasing. However, if specification~(\ref%
{quantile-spec}) is incorrect, there is no guarantee that $\tau \rightarrow
\beta \left( \tau \right) ^{\top }x$ should be monotone. This phenomenon has
been widely recognized in the literature on quantile regression and is known
as the \emph{quantile crossing problem}. To address the quantile crossing
problem, one idea may be to impose directly the monotonicity of $\tau
\rightarrow \beta \left( \tau \right) ^{\top }x$ as an additional constraint
in problem~(\ref{qreg-pop-integrated}). This has been done by Koenker and Ng
(\citeyear{koenker_inequality_2005}) but remains computationally challenging and the interpretation of the
result is not obvious.

A more indirect approach consists of the following. Rather than imposing the
monotonicity of $\tau \rightarrow \beta \left( \tau \right) ^{\top }x$, one
can impose the (weaker) constraint that $\tau \rightarrow 1\left\{ y\geq
\beta \left( \tau \right) ^{\top }x\right\} $ should be nonincreasing in $%
\tau $. Consider the problem%
\begin{eqnarray}
\min_{\beta \in \mathbb{R}^{k\times \left[ 0,1\right] }} &&\mathbb{E}_{_{%
\mathbf{P}}}\left[ \int_{0}^{1}\rho _{\tau }\left( Y-\beta \left( \tau
\right) ^{\top }X\right) d\tau \right]  \label{qreg-pop-constrained} \\
s.t.~ &&1\left\{ y\geq \beta \left( \tau \right) ^{\top }x\right\} \geq
1\left\{ y\geq \beta \left( \tau ^{\prime }\right) ^{\top }x\right\}
~\forall \tau \leq \tau ^{\prime },\forall x\in \mathbb{R}^{k},y\in \mathbb{R%
}  \notag
\end{eqnarray}

The solution to the previous problem now has a very straightforward
interpretation.

\begin{theorem}[Carlier-Chernozhukov--De Bie-Galichon]
\label{thm:ccdg}If the map $\beta $ is solution to problem~(\ref{qreg-pop}),
then denoting $b\left( \tau \right) =\int_{0}^{\tau }\beta \left( t\right)
dt $, and letting%
\begin{equation}
\psi \left( x,y\right) =\max_{\tau \in \left[ 0,1\right] }\left\{ \tau
y-x^{\top }b\left( \tau \right) \right\} ,  \label{psi-from-b}
\end{equation}%
the pair $\left( b,\psi \right) $ will be solution to the following problem 
\begin{align}
\max_{b,\psi }& \mathbb{E}_{_{\mathbf{P}}}\left[ X\right] ^{\top
}\int_{0}^{1}b\left( \tau \right) d\tau +\mathbb{E}_{_{\mathbf{P}}}\left[
\psi \left( X,Y\right) \right]  \label{vqr-dual} \\
s.t.~& x^{\top }b\left( \tau \right) +\psi \left( x,y\right) \geq \tau
y.~\blacksquare  \notag
\end{align}

Conversely, if $\left( b,\psi \right) $ is solution to problem~(\ref%
{vqr-dual}) and if $b$ is differentiable, then $\beta \left( \tau \right)
=b^{\prime }\left( \tau \right) $ is a solution to problem~(\ref{qreg-pop}).
\end{theorem}

\bigskip

Theorem~\ref{thm:ccdg} sheds new insights on quantile regression. Indeed, an
extension of Monge-Kantorovich duality worked out in Carlier, Chernozhukov
and\ Galichon (\citeyear{carlier_vector_2016}) shows that problem~(\ref{vqr-dual}) is the dual problem
to 
\begin{eqnarray}
\min_{\pi \in \mathcal{M}\left( \mathcal{U}\left( \left[ 0,1\right] \right) ,%
\mathbf{P}\right) } &&\mathbb{E}_{\pi }\left[ (Y-U)^{2}\right]
\label{vqr-primal} \\
s.t. &&\mathbb{E}\left[ X|U\right] =\mathbb{E}\left[ X\right]  \notag
\end{eqnarray}%
where $\pi \in \mathcal{M}\left( \mathcal{U}\left( \left[ 0,1\right] \right)
,\mathbf{P}\right) $ means that if $\left( U,X,Y\right) \sim \pi $, then $%
U\sim \mathcal{U}\left( \left[ 0,1\right] \right) $ and $\left( X,Y\right)
\sim \mathbf{P}$. If $\left( b,\psi \right) $ is a solution to~(\ref%
{vqr-dual}) with $b$ differentiable and $\left( U,X,Y\right) \sim \pi $ is a
solution to~(\ref{vqr-primal}), then letting $\beta \left( \tau \right)
=b^{\prime }\left( \tau \right) $, one has the representation%
\begin{equation}
Y=X^{\top }\beta \left( U\right)  \label{vqr-factor-rep}
\end{equation}%
where $X$ is mean-independent from $U$. \ Beyond the case when $Y$ is
scalar, this formulation allows to get a multivariate extension of quantile
regression using the notion of vector quantiles; see Carlier, Chernozhukov,
Galichon (\citeyear{carlier_vector_2016}).

\subsection{Partial identification and random sets}

Some problems in econometrics specify incomplete restrictions between a
model and an observed variable. Assume, following Galichon and Henry (\citeyear{galichon2011set}),
that we observe a random variable $Y\sim Q$ valued in $\mathcal{Y}$, and
that the restrictions given by the model specify $Y\in \Gamma _{\theta
}\left( X\right) $, and $X\sim P$, where $X$ is a data-generating process
valued in $\mathcal{X}$ and $\theta \in \Theta $ is a parameter of the
model. $\Gamma _{\theta }$ is a correspondence from $\mathcal{X}$ to $%
\mathcal{Y}$, such that $\Gamma _{\theta }\left( x\right) $ is a subset of $%
\mathcal{Y}$. The \emph{identified set} is the set $\Theta _{I}$ of $\theta
\in \Theta $ such that there is a joint distribution $\pi \in \mathcal{M}%
\left( P,Q\right) $ with $\mathbb{E}_{\pi }\left[ 1\left\{ Y\notin \Gamma
_{\theta }\left( X\right) \right\} \right] =0$. Such a problem can be recast
as an optimal transport problem 
\begin{equation}
V=\min_{\pi \in \mathcal{M}\left( P,Q\right) }\mathbb{E}_{\pi }\left[
1\left\{ Y\notin \Gamma _{\theta }\left( X\right) \right\} \right]
\label{set-inclusion}
\end{equation}%
By working on the dual of this problem, one obtains Strassen's theorem
(Strassen, \citeyear{strassen_existence_1965})%
\begin{equation}
V=\max_{B}\left\{ Q\left( B\right) -P\left( \Gamma ^{-1}\left( B\right)
\right) \right\} ,  \label{set-inclusion-dual}
\end{equation}%
where the maximum extends over the Borel sets $B$ of $\mathcal{Y}$.
Therefore $\theta \in \Theta _{I}$ if and only if $Q\left( B\right) \leq
P\left( \Gamma ^{-1}\left( B\right) \right) $ for all $B$. \ The sample
version of problem~(\ref{set-inclusion}) allows to use optimal assignment
algorithms as efficient computational tools to decide if $\theta \in \Theta
_{I}$, and dual formulation~(\ref{set-inclusion-dual}) allows to do
inference (Galichon and Henry, \citeyear{galichon_test_2009}).

\subsection{Generalized linear models}

Consider a generalized linear model (GLM)\ with 2-way fixed effects. The
observations are $ij$; the dependent variable is $\hat{\pi}_{ij}$, while the
explanatory variables are $\Phi =\left( \phi _{ij}^{k}\right) _{ij,k}$ for $%
k\in \left\{ 1,...,K\right\} $ and the $i$ and $j$ fixed effects. If $l$ is
the link function, which is increasing and continuous, the model is written
as%
\begin{equation}
\mathbb{E}\left[ \hat{\pi}_{ij}|\phi _{ij}^{k},i,j\right] =l^{-1}\left(
\left( \Phi \beta \right) _{ij}-u_{i}-v_{j}\right) .  \label{glm-2way}
\end{equation}%
Denote $p_{i}=\sum_{j}\hat{\pi}_{ij}$ and $q_{j}=\sum_{i}\hat{\pi}_{ij}$ the
margins of $\hat{\pi}$. Letting $L$ be a primitive of $l$, and letting $%
L^{\ast }\left( w\right) =\max_{z}\left\{ wz-L\left( z\right) \right\} $ be
its convex conjugate, which is a primitive of $l^{-1}$, one can show that
the GLM model can be fit using%
\begin{equation}
\min_{\beta }\left\{ W\left( \beta \right) -\sum_{ij}\hat{\pi}_{ij}\left(
\Phi \beta \right) _{ij}\right\}  \label{glm-fit}
\end{equation}%
where%
\begin{eqnarray}
W\left( \beta \right) =\max_{\pi _{ij}\geq 0} &&\left\{ \sum_{ij}\pi
_{ij}\left( \Phi \beta \right) _{ij}-\sum_{ij}L\left( \pi _{ij}\right)
\right\}  \label{ot-general-regul-primal} \\
s.t.~ &&\sum_{j}\pi _{ij}=p_{i},~\sum_{i}\pi _{ij}=q_{j}  \notag
\end{eqnarray}%
is a regularized optimal transport problem which can be equivalently
expressed by its dual:%
\begin{equation}
W\left( \beta \right) =\min_{u_{i},v_{j}}\left\{
\sum_{i}p_{i}u_{i}+\sum_{j}q_{j}v_{j}+\sum_{ij}L^{\ast }\left( \left( \Phi
\beta \right) _{ij}-u_{i}-v_{j}\right) \right\} .
\label{ot-general-regul-dual}
\end{equation}

In particular, the $\log $ link function $l\left( z\right) =\ln z$ yields $%
l^{-1}\left( t\right) =\exp \left( t\right) $, and thus $L\left( z\right)
=z\left( \ln z-1\right) $, and $L^{\ast }\left( t\right) =\exp \left(
t\right) $, and $W\left( \beta \right) $ is the solution to an entropy
regularized optimal transport problem, as described in paragraph~\ref%
{par:regul-ot}.

\subsection{Hide-and-seek games}

In \citeyear{von20161}, von Neumann described the following two-person, zero-sum game. Let $%
\left( K_{ij}\right) $ be a $N\times N$ matrix with positive terms. There
are two players, \textquotedblleft Hider\textquotedblright\ and
\textquotedblleft Seeker\textquotedblright . Hider plays first and hides in
a cell $\left( i,j\right) $. \ Playing second, Seeker highlights either a
row or a column they claims contains Hider. If Seeker's claim is correct, then Hider
pays Seeker $K_{ij}>0$, otherwise 0. 

Hider's mixed strategy is described
by a vector of probabilities  $\pi _{ij}$ of hiding in cell $ij$. Once Hider has played, Seeker picks
either a column $i^{\prime }$ or a column $j^{\prime }$, whichever of these maximizes
$\sum_{j^{\prime }}K_{ij^{\prime }}\pi _{ij^{\prime }}\,$over $i$\ and $%
\sum_{i^{\prime }}K_{i^{\prime }j}\pi _{i^{\prime }j}$ over $j$. Let us
denote $\left( a,b\right) $ the vector of mixed strategies of Seeker, where $%
a_{i}\geq 0$ is the probability of highlighting a row $i$, and $b_{j}\geq 0$
is the probability of highlighting a column $j$, and $\sum_{i=1}^{n}a_{i}+%
\sum_{j=1}^{n}b_{j}=1$. If Hider plays strategy $\pi $ and Seeker plays
strategy $\left( a,b\right) $, the expected payoff of Seeker is therefore
\begin{equation*}
\sum_{ij}\left( a_{i}+b_{j}\right) K_{ij}\pi _{ij}
\end{equation*}
and hence the value of this zero-sum game for Seeker is obtained by minimizing the
above expression over $x_{ij}\geq 0$, $\sum_{ij}x_{ij}=1$, and maximizing it
over $\left( a,b\right) \geq 0$ such that $\sum_{i}a_{i}+\sum_{j}b_{j}=1$.

Von Neumann showed that this game is intimately connected with an optimal
transport problem. Indeed,
\begin{equation*}
\begin{array}{llll}
V^{-1}= & \max_{\pi \geq 0}\sum_{ij}\pi _{ij} K_{ij}^{-1}  & = &
\min_{u\geq 0,v\geq 0}\sum_{i}\frac{u_{i}}{n}+\sum \frac{v_{j}}{n} \\
& s.t.~\left\{
\begin{array}{l}
\sum_{j}\pi _{ij}\leq 1/n \\
\sum_{i}\pi _{ij}\leq 1/n%
\end{array}%
\right.  &  & s.t.~u_{i}+v_{j}\geq  K_{ij}^{-1} %
\end{array}%
\end{equation*}%
and the solution $\pi _{ij}\geq 0$ to the primal problem yields Hider's
optimal strategy, while setting $a_{i}=V u_{i}/n$ and $b_{j}=V v_{j}/n$ yields
Seeker's optimal strategy.

Although von Neumann's paper appeared in \citeyear{von20161}, it seems that this important connection between a zero-sum game and a linear programming problem was known to him decades earlier, in anticipation of Dantzig's general connection between linear programming and zero-sum games, cf.~Dantzig
(\citeyear{dantzig1951proof}). See a historical perspective in Kuhn and Tucker (\citeyear{kuhn1958john}).

\section{The mathematics of optimal transport}

\subsection{Network formulation\label{par:network}}

As explained in Galichon (\citeyear{galichon_optimal_2016}), chapter 8, the optimal transport problem
has the structure of a \emph{min-cost flow} problem. Introduce a network
whose set of nodes is $\mathcal{Z}=\mathcal{X}\cup \mathcal{Y}$ and whose
set of arcs is $\mathcal{A}=\mathcal{X}\times \mathcal{Y}$. Such a network
is called a \emph{bipartite} one. Define an $\mathcal{A\times Z}$ matrix $%
\nabla $ which is such that $\nabla _{xy,z}=1\left\{ z=y\right\} -1\left\{
z=x\right\} $. Consider the \textquotedblleft change of sign
trick\textquotedblright\ where one defined $\tilde{q}=\left( -p^{\top
},q^{\top }\right) ^{\top }$ and $\tilde{v}=\left( -u^{\top },v^{\top
}\right) ^{\top }$. Define $c_{xy}=-\Phi _{xy}$. The vector $\tilde{q}$
should be interpreted as a vector of quantities, while the vector $\tilde{v}$
should be interpreted as a vector of prices.

Call $C\left( \tilde{q}\right) $ the value of the optimal transport problem,
which rewrites under its primal form as 
\begin{eqnarray}
C\left( \tilde{q}\right) =\min_{\pi \geq 0} &&\pi ^{\top }c  \label{C} \\
s.t.~ &\nabla ^{\top }\pi =&\tilde{q}  \notag
\end{eqnarray}%
when $q^{\top }1_{\mathcal{Z}}=0$, and $C\left( \tilde{q}\right) =+\infty $
otherwise. Equivalently, $C\left( \tilde{q}\right) $ can be expressed by its
dual value as%
\begin{eqnarray*}
C\left( \tilde{q}\right) =\max_{\tilde{v}} &&\tilde{q}^{\top }\tilde{v} \\
s.t.~ &\nabla \tilde{v}\leq &c.
\end{eqnarray*}

This is an instance of the \emph{min-cost flow problem}, which makes sense
more generally on any (not necessarily bipartite)\ network.

\subsection{Equilibrium expression}

By convex duality, denoting 
\begin{equation}
C^{\ast }\left( \tilde{v}\right) =\max_{\tilde{q}}\left\{ \tilde{q}^{\top }%
\tilde{v}-C\left( \tilde{q}\right) \right\} ,  \label{Cstar}
\end{equation}
the convex conjugate of $C$, it can be seen that $C^{\ast }\left( \tilde{v}%
\right) =0$ if $\nabla \tilde{v}\leq c$ and $+\infty $ otherwise, and one
has 
\begin{equation}
C\left( \tilde{q}\right) =\max_{\tilde{v}}\left\{ \tilde{q}^{\top }\tilde{v}%
-C^{\ast }\left( \tilde{v}\right) \right\} .  \label{CfromCstar}
\end{equation}

Further, the set of $\tilde{v}=\left( -u,v\right) $ where $u$ and $v$ are
solutions to problem~(\ref{dual-noregul}) are the maximizers of~(\ref%
{CfromCstar}).

In the case of the entropy regularized problem~(\ref{primal-regul})-(\ref%
{dual-regul}), these expressions become respectively%
\begin{eqnarray}
C_{\sigma }\left( \tilde{q}\right) =\min_{\pi \geq 0} &&\pi ^{\top }c+\sigma
\pi ^{\top }\log \pi  \label{C-regul} \\
s.t.~ &\nabla ^{\top }\pi =&\tilde{q}  \notag
\end{eqnarray}%
and%
\begin{equation}
C_{\sigma }^{\ast }\left( \tilde{v}\right) =\sigma 1^{\top }\exp \left( 
\frac{c-\nabla \tilde{v}}{\sigma }\right) .  \label{Cstar-regul}
\end{equation}

Keeping in mind the interpretation of $\tilde{q}$ as quantities and $\tilde{v%
}$ as prices, one should view $C\left( \tilde{q}\right) $ as a cost
function, \ expression $\tilde{v}^{\top }\tilde{q}-C_{\sigma }\left( \tilde{q%
}\right) $ as a profit, and $C^{\ast }\left( \tilde{v}\right) $ as an
indirect profit function. Hence, expression~(\ref{Cstar}) should be viewed
as a profit maximization problem. The optimal transport problem consists of
looking for the potentials $\tilde{v}$ that maximize $\tilde{q}^{\top }%
\tilde{v}-C^{\ast }\left( \tilde{v}\right) $. By convex duality (see chapter
6 of Galichon, \citeyear{galichon_optimal_2016}), this is equivalent with 
\begin{equation}
\tilde{v}\in \partial C\left( \tilde{q}\right) ,  \label{inv-demand}
\end{equation}%
which, still by convex analysis, is equivalent with 
\begin{equation}
\tilde{q}\in \partial C^{\ast }\left( \tilde{v}\right) .  \label{demand}
\end{equation}%
Therefore, $\partial C^{\ast }$ should be interpreted as a \emph{supply
correspondence}, while $\partial C=\left( \partial C^{\ast }\right) ^{-1}$
should be interpreted as an \emph{inverse supply correspondence}. The same
intepretation extends immediately to the regularized versions of these
objects.

\subsection{Mathematical structures}

The optimal transport problem is blessed with the priviledge to belong to
the intersection of two rich theories: convex optimization and gross
substitutes. There are, broadly speaking, two structures whithin which the
equilibrium problem%
\begin{equation}
\tilde{q}\in Q\left( \tilde{v}\right)  \label{equil-pb}
\end{equation}%
is well understood.

\begin{itemize}
\item The first one is \emph{convex optimization}: $Q$ is the
subdifferential of a convex function. Then the problem is a convex
optimization problem, and convex optimization can be put to use to solve
problem~(\ref{equil-pb}).

\item The second case is \emph{gross substitutes}: loosely speaking, $q_{x}$
cannot increase when $\tilde{v}_{y}$ increases ($x\neq y$). This setting is
needed for coordinate update algorithms such as Jacobi or Gauss-Seidel to
converge, see Rheinboldt (\citeyear{ortega_iterative_2000}).
\end{itemize}

In optimal transport, both structures are met, as we shall now see.

\subsubsection{Convex optimization}

Recall that the cost function $C$ and the indirect cost function $C^{\ast }$
defined above are convex functions, which are dual one to another in the
sense of convex analysis. It follows that $\partial C^{\ast }\left( \tilde{v}%
\right) $ and $\partial C\left( \tilde{q}\right) $ are convex sets, and
problems~(\ref{inv-demand}) and~(\ref{demand}) can be solved as convex
optimization problems dual to each other, respectively~(\ref{CfromCstar})
and~(\ref{Cstar}). In the unregularized case, these problems are linear
programming problems. In the regularized case, the convexity structure is
retained, but the problems are of course no longer linear.

\subsubsection{Gross substitutes\label{par:gs}}

When $\sigma >0$ it is easy to see that the indirect profit function $%
C_{\sigma }^{\ast }\left( \tilde{v}\right) $ is submodular. It is not very
hard to extend this result to the unregularized case to show that $C^{\ast
}\left( \tilde{v}\right) $ is submodular as well. \ As a result, the
corresponding supply function satisfies Kelso and Crawford's (\citeyear{kelso_job_1982}) \emph{%
gross substitutes} property.

A remark is in order here. It may be a surprise that the optimal transport
problem has the gross substitutes property, as common sense suggests that
workers and firms should be \emph{complements}, and not substitutes.
However, keep in mind the \textquotedblleft change-of-sign
trick\textquotedblright\ implemented at paragraph~\ref{par:network}: we
defined $\tilde{v}=\left( -u^{\top },v^{\top }\right) ^{\top }$, and
therefore we switched the sign of the worker's payoffs (and of their
quantities accordingly). This change of sign is the reason why the optimal
transport problem, in spite of being a problem with complementarities,
reformulates as a problem with gross substitutes. See Sun and Yang (\citeyear{sun2006equilibria}).

We can formulate gross substitutes properties of $C$ and $C^{\ast }$ in the
language of L- and M-convexity, introduced by Murota (\citeyear{murota1998discrete}). Indeed, as the
domain of $C$ is the set of $\tilde{q}$ such that $\tilde{q}^{\top }1_{%
\mathcal{Z}}=0$, and as $C^{\ast }\left( \tilde{v}+\lambda 1_{\mathcal{Z}%
}\right) =C^{\ast }\left( \tilde{v}\right) $ for all $\lambda \in \mathbb{R}$%
, it follows that $C$ is a \emph{M-convex function} and $C^{\ast }$ is a 
\emph{L-convex function}, and the supply bundle $\partial C^{\ast }\left( 
\tilde{v}\right) $ is a\emph{\ M-convex set} of $\mathbb{R}^{\mathcal{Z}}$,
while $\partial C\left( \tilde{q}\right) $ is a \emph{L-convex set}, still
in the terminology of the same author. In particular, $\partial C\left( 
\tilde{q}\right) $ is a \emph{lattice}, while $\partial C^{\ast }\left( 
\tilde{v}\right) $ is a \emph{base polyhedron}.

\subsection{Extensions}

As we have seen just above, the optimal transport problem can be formulated (at least under its regularized form) as a set of nonlinear equations $Q(p) = q$, where $Q$ happens to be the subdifferential
of a convex function which is also submodular, and hence optimal transport
belongs to both convexity \emph{and} gross substitutes families. Some extensions of the
optimal transport problem retain both convexity and gross substitutes. This
is the case of the min-cost flow problem, for instance, as described in
paragraph~\ref{par:network}. \ 

\subsubsection{Problems that retain convexity, but not substitutability}

Some problems retain convex optimization but not gross substitutes, such as
one-to-many matching problems with transferable utility, see a related
discussion in Azevedo and Hatfield (\citeyear{azevedo_existence_2018}). Vector quantile regression, discussed above in paragraph~\ref{par:quantile-regression}, falls in that category, too.

In ongoing work with Pauline Corblet and Jeremy Fox~\citeyear{fox_dynamic_2020}, we investigate a problem of dynamic matching that retains most of the convexity structure of optimal transport. The problem we study is a two-sided version of Rust~(\citeyear{rust1987optimal})'s model. More specifically, assume that conditional of a worker of type $X=x$ matching with a firm of type $y$, there is a probability $\mathbb{P}_{x^\prime|xy}$ that the worker will transition to type $x^\prime$ at the next period, and a probability  $\mathbb{Q}_{y^\prime|xy}$ that the firm will transition to type $y^\prime$. In this case, the joint matching surplus should be the sum of the short-term surplus $\Phi_{xy}$ and the expected discounted future payoffs of the worker and of the firm, respectively denoted $\beta \mathbb{P}\left[ u_{X}|xy\right]$ and $\beta \mathbb{Q}\left[ v_{Y}|xy\right]$.

In this case, when $\beta = 1$, Corblet et al.~(\citeyear{fox_dynamic_2020}) show that both the equilibrium computation and the estimation can be handled by the following saddle-point problem:
\begin{equation*}
\max_{n,m}\min_{u,v,\lambda }H\left( n,m,u,v,\lambda \right)
\end{equation*}
where one has defined $H\left( n,m,u,v,\lambda \right) = $
\begin{align*}  \left\{
\begin{array}{l}
2\sum_{xy\in \mathcal{X}\times \mathcal{Y}}\sqrt{n_{x}m_{y}}\exp \left(
\frac{\sum_{k}\phi _{xy}^{k}\lambda _{k}+\mathbb{P}\left[ u_{X}|xy%
\right] +\mathbb{Q}\left[ v_{Y}|xy\right] -u_{x}-v_{y}}{2}\right)
\\
+\sum_{x\in \mathcal{X}}n_{x}\exp \left( \sum_{k}\phi _{x0}^{k}\lambda _{k}+%
\mathbb{E}\left[ u_{X^{\prime }}|X=x\right] -u_{x}\right) \\
+\sum_{y\in
\mathcal{Y}}m_{y}\exp \left( \sum_{k}\phi _{0y}^{k}\lambda _{k}+\mathbb{E}%
\left[ v_{Y^{\prime }}|Y=y\right] -v_{y}\right)  \\
-\sum_{x\in \mathcal{X}}n_{x}-\sum_{y\in \mathcal{Y}}m_{y}%
\end{array}%
\right.
\end{align*}
which is convex in $(n,m)$ and concave in $(u,v,\lambda)$. Corblet et al.~(\citeyear{fox_dynamic_2020}) use this formulation to derive an algorithm to estimate the structural parameter $\lambda$ efficiently. They find that the algorithm extends to the case $\beta <1$.
Dupuy et al.~(\citeyear{ciscato_2020}) apply these ideas to family economics and fertility decisions. 
 
\subsubsection{Problems that retain substitutability, but not convexity}
On the contrary, some problems retain the gross substitutes property, but
not the convexity one. This is the case with one-to-one matching
models with nontransferable utility, as shown by Adachi (\citeyear{adachi_characterization_2000}), and with
one-to-one matching models with imperfectly transferable utility, to handle
in particular taxes, salary caps, public goods, etc. See Galichon, Kominers
and Weber (\citeyear{galichon_costly_2019}). Non-additive random utility models and hedonic models beyond
quasi-linear utility are also in this case. To handle these challenges, a more
general framework is needed, the \emph{equilibrium flow problem}, which is
the subject of current ongoing work by the author with Larry Samuelson and
Lucas Vernet~(\citeyear{gsv_equilibrium_2021}). The equilibrium flow problem posits three objects. First a network  $(\mathcal{Z},\mathcal{A})$ is defined as in paragraph~\ref{par:network}, where $xy \in \mathcal{Z}$ is interpreted as the existence of a trade route from node $x$ to node $y$, and whose node-incidence matrix is denoted $\nabla$. Second, a vector of outflows $q \in \mathbb{R}^\mathcal{Z}$, where $q_z$ is interpreted as the mass that must leave the network at $z$ ($q_z < 0$ means that mass actually appears at $z$). One assumes  that $\sum_{z \in \mathcal{Z}} q_z = 0$, so all the mass that enters the networks must leave it. Finally, a set of \emph{connection functions}
 $G_{xy}: \mathbb{R} \rightarrow \mathbb{R}$ for each $xy \in \mathcal{A} $, which are increasing and whose interpretation is that
$ G_{xy}(p_y) - p_x $ is the profit of a \emph{carry trade}, consisting of purchasing one unit of the commodity at price $p_x$ at node $x$, shipping to $y$, and selling at price $p_y$ at node $y$.

Given these inputs, the equilibrium flow problem consists of determining a vector of flows $\mu \in \mathbb{R}_+^\mathcal{Z}$ and prices $p \in \mathbb{R}^\mathcal{Z}$ such that:

(i) \emph{mass balance} holds: the sum of mass that arrives at $z$ minus the sum that leaves is equal to $q_z$, that is, $ \nabla^\top \mu = q $. 

(ii) \emph{absence of arbitrage} holds: there cannot be a positive rent associated with the carry trade over any arc, that is, 
$p_x \geq G_{xy}(p_y)$ for any $xy \in \mathcal{A}$.

(iii) \emph{individual rationality} holds: if the carry trade over arc $xy$ is actually performed, then the associated profit cannot be negative, and thus,
$\mu_{xy}>0$ implies $p_x = G_{xy}(p_y).$ 

Galichon, Samuelson and Vernet~(\citeyear{gsv_equilibrium_2021}) show that this framework is general enough to encompass optimal transport problems, min-cost flow problems including shortest path problems, matching models with imperfectly transferable utility, hedonic models, and supply chain problems.

\section{Concluding discussion}

To conclude, an attempt should be made to explain the claim to
\textquotedblleft unreasonable effectiveness\textquotedblright\ of optimal
transport in economics, alluding to a celebrated formula of Wigner (\citeyear{wigner_unreasonable_1960}).
We believe that one of the reasons of the prevalence of optimal transport in
economics is that the former strikes a good compromise between what models
would like to capture and what they are capable of capturing.

Economics is, in a broad sense, the study of complementarities: capital and
labor, worker and firms, supply and demand, buyers and sellers... all exhibit
some complementarity which is at the source of economic activity. However,
as it is now well understood since the insights of Kelso and Crawford (\citeyear{kelso_job_1982}), problems with complementarities are hard to handle, and in
particular, hard to compute. Fortunately, due to the bipartite structure,
the \textquotedblleft change-of-sign trick\textquotedblright\ described in
paragraph~\ref{par:network} allowed us to reformulate the problem as a
problem with gross substitutes, and therefore, let us enjoy the
computational and structural benefits of a problem with gross substitutes.
In some sense, the bipartite structure of optimal transport is a meeting
point between the complementarity that models would like to capture, and the
substitutability structure that they they are able to capture.

To make another analogy with Physics, the situation is similar to the
two-body problem in cosmology, which has a
tractable formulation and can be fully worked out -- while the $n$-body
problem with $n$ larger than two is notoriously hard. Fortunately, just as
in cosmology where many situations can be satisfactorily approximated by a
two-body problem, in economics, many phenomenons can be captured using the
bipartite approximation. We have surveyed some of these applications in the
present paper, but certainly not in an exhaustive way. And optimal transport is a galaxy where there are many
more planets, only waiting to be explored.

\printbibliography

\end{document}